# Intrinsic ferroelectricity in Y-doped HfO$_2$ thin films

Yu Yun,[1,†] Pratyush Buragohain,[1,†] Ming Li,[1,†] Zahra Ahmadi,[2] Yizhi Zhang,[3] Xin Li,[1] Haohan Wang,[1] Lingling Tao,[1] Haiyan Wang,[3] Jeffrey E. Shield,[2,4] Evgeny Y. Tsymbal,[1,4]* Alexei Gruverman,[1,4]* Xiaoshan Xu[1,4]*

[1] Department of Physics and Astronomy, University of Nebraska-Lincoln, Lincoln, Nebraska 68588, USA
[2] Department of Mechanical and Materials Engineering, University of Nebraska–Lincoln, Lincoln, Nebraska 68588, USA
[3] School of Materials Engineering, Purdue University, West Lafayette, Indiana 47907, USA
[4] Nebraska Center for Materials and Nanoscience, University of Nebraska, Lincoln, Nebraska 68588, USA

†Joint first authors.

*Corresponding authors: tsymbal@unl.edu, agruverman2@unl.edu, xiaoshan.xu@unl.edu

## Abstract

Ferroelectric HfO$_2$-based materials hold great potential for widespread integration of ferroelectricity into modern electronics due to their robust ferroelectric properties at the nanoscale and compatibility with the existing Si technology. Earlier work indicated that the nanometer crystal grain size was crucial for stabilization of the ferroelectric phase of hafnia. This constraint caused high density of unavoidable structural defects of the HfO$_2$-based ferroelectrics, obscuring the intrinsic ferroelectricity inherited from the crystal space group of bulk HfO$_2$. Here, we demonstrate the intrinsic ferroelectricity in Y-doped HfO$_2$ films of high crystallinity. Contrary to the common expectation, we show that in the 5% Y-doped HfO$_2$ epitaxial thin films, high crystallinity enhances the spontaneous polarization up to a record-high 50 μC/cm$^2$ value at room temperature. The high spontaneous polarization persists at reduced temperature, with polarization values consistent with our theoretical predictions, indicating the dominant contribution from the intrinsic ferroelectricity. The crystal structure of these films reveals the Pca2$_1$ orthorhombic phase with a small rhombohedral distortion, underlining the role of the anisotropic stress and strain. These results open a pathway to controlling the intrinsic ferroelectricity in the HfO$_2$-based materials and optimizing their performance in applications.

# Introduction

Ferroelectric materials exhibit switchable spontaneous electric polarization, which makes them promising for application in modern electronics, especially for information storage and processing [1]. Conventional $ABO_3$ perovskite ferroelectrics suffer from poor scalability due to the increasing depolarization field at reduced thicknesses and incompatibility with the current Si-technology [2-5]. The recent discovery of robust ferroelectricity at the nanoscale in hafnium oxide ($HfO_2$) based materials, which have long been used as high-k dielectrics, makes the widespread integration of ferroelectricity into nanoscale electronics feasible.[6,7]

Earlier reports suggested that to establish ferroelectricity, it is crucial for the $HfO_2$-based materials to consist of nanometer-sized crystal grains, because the ferroelectric orthorhombic structural phase ($Pca2_1$, $o$-phase, Fig. 1a) is unstable in ambient conditions. All the stable structural phases of $HfO_2$, i.e., the monoclinic phase ($P2_1/c$, $m$-phase) at room temperature, the tetragonal phase ($P4_2/mnc$, $t$-phase) above 2100 K, and the cubic phase (Fm-3m, $c$-phase) above 2800 K [8] are not ferroelectric. The small crystal grain sizes lead to proliferation of structural defects (low crystallinity) in the form of grain boundaries, which are expected to undermine and obscure ferroelectric properties.[9] Indeed, it has been challenging to elucidate the crystal structure of the ferroelectric $o$-phase; to date, most of the experimentally observed spontaneous polarization values are significantly lower than the theoretically predicted ones (40-60 $\mu C/cm^2$) [10-14].

Being able to fabricate $HfO_2$-based ferroelectrics with the minimally possible defect density, or high crystallinity, will not only allow elucidation of intrinsic ferroelectric properties, but also enable better performance in device applications. The key is to enhance the stability of the ferroelectric $o$-phase, which is typically achieved by constraining the grain size making use of the two main mechanisms. (1) The small grains size enhances the stability of the $o$-phase because the surface energy of the $o$-phase is larger than that of the $m$-phase.[10,15-19] (2) $HfO_2$-based ferroelectrics are typically obtained by fast cooling from the high-temperature $t$-phase taking advantage of the smaller energy of the $o/t$ interface compared with that of the $m/t$ interface, even though the bulk $m$-phase is more stable; the size of the crystal grains is then limited by the rapid cooling process.[17,19-30] On the other hand, when the stability of the $o$-phase is enhanced by additional mechanisms, such as doping and epitaxial growth, the ferroelectricity may reconcile with high crystallinity in $HfO_2$-based materials. Among various dopants of $HfO_2$ [6,15,29,31-34], yttrium stands out for stabilizing the $o$-phase $HfO_2$ in thick films [35,36] and even in bulk [20]. In addition, in ultra-thin epitaxial films, the large surface area and the anisotropic stress and strain have been suggested to enhance the stability of the $o$-phase [10,15-19].

In this work, we challenge the common belief that the smaller grain size is required to stabilize the ferroelectric $o$-phase in $HfO_2$-based thin films. We investigate molar 5% $YO_{1.5}$ doped $HfO_2$ (YHO) films, focusing on high-temperature epitaxial growth on LSMO (001) / STO (001) and LSMO (110) /STO (110) substrates, where LSMO and STO stand for $La_{0.7}Sr_{0.3}MnO_3$ and $SrTiO_3$, respectively. We demonstrate that higher crystallinity of the films actually enhances spontaneous polarization up to a record-high value of 50 $\mu C/cm^2$ at room temperature. We show that the high polarization only moderately decreases at low temperature suggesting the dominant contribution from the intrinsic ferroelectricity.

## Results and Discussion
**Positive correlation between $P_r$ and crystallinity**

A typical $\theta$-$2\theta$ x-ray diffraction (XRD) scan for YHO (111) /LSMO (001) thin films grown at substrate temperature $T_s$ = 890 °C in $O_2$ pressure of 70 mTorr (optimal condition for spontaneous polarization, same below) is shown in Fig. 1b. The clear Laue oscillations around the YHO peak indicate smooth surface and interfaces. The thickness of the LSMO and YHO layers are ≈ 25 nm and ≈ 10 nm, respectively, according to x-ray reflection (XRR) (Supplementary Fig. S1). The peak at $2\theta \approx 30°$ can be assigned to the diffraction of the pseudo cubic $(111)_{pc}$ plane [24-28,37].

Fig. 1c shows the remanent Polarization-Voltage (P-V) loop measured using the positive up and negative down (PUND) method at room temperature for the YHO films grown at optimal condition. The spontaneous or remanent polarization ($P_r$) value is approximately 36 μC/cm$^2$, which is larger than all polarization values reported for Y-doped $HfO_2$ films.[20-22,35,36,38] The coercive voltage ($E_c$) is much larger than $E_c$ of the polycrystalline films [6,15,16,20,35] and slightly larger than $E_c$ of the 5-nm-thick HZO (111) /LSMO (001) films.[24]

The in-plane size of the crystal grains was estimated using x-ray diffraction rocking curves, which measures the size of reciprocal lattice point along the in-plane direction and in turn the in-plane size of the crystallites for epitaxial films. As shown Fig. 1b inset for the films grown at optimized condition (see also Fig. S2), the rocking curves clearly consist of a narrow peak sitting on a broad peak, corresponding to small (≈ 10 nm) and large (≈ 100 nm) crystal grains respectively, according to the Scherrer formula [39]. The large grain is an order of magnitude larger than the typical grain size (< 10 nm) found in polycrystalline films [16,17] and previously reported epitaxial films.[24,28]

To elucidate whether the high crystallinity contributes to the large polarization, we studied YHO films with different growth temperature $T_s$, which is expected to be critical for the microstructure of the $HfO_2$-based films with multiple competing phases.[27] The $O_2$ pressure is fixed at 70 mTorr, similar to the optimal oxygen pressure (0.1 mbar) in the growth of HZO on LSMO (001) [24,27]. At lower temperature ($\leq$ 800°C), the films contain traces of m-phase as indicated by the $(-111)_m$ peak at $2\theta \approx 28.5°$ (Supplementary Fig. S3), which disappears at higher temperature ($\geq$ 850°C).

As temperature changes, the rocking curves (Fig. S2) remain comprised of the narrow peak and the broad peak while their relative weight changes dramatically, corresponding to a large change of the volume fraction of the small and large grains. Since more large grains means less grain boundary defects, here we define the large/small grain ratio (area ratio of the narrow and broad peaks) as a quantitative measure of the crystallinity.

The $T_s$ dependence of the crystallinity (large/small grain ratio) shows a non-monotonic trend, with a peak at $T_s$ = 890°C, as shown in Fig. 1d. Below 890°C, the crystallinity increases with temperature, most probably due to the reduction of the m-phase. Above 890°C, the crystallinity decreases with temperature, likely due to the decay of the LSMO layer (Supplementary Fig. S4). The $P_r$ as a function of $T_s$ is also plotted in Fig. 1d, which shows a non-monotonic trend in perfect match with the $T_s$-dependence of crystallinity (large/small grain ratio).

In contrast, the YHO (111) peak shifts monotonically to higher $2\theta$ at higher temperature, indicating reduced $d_{(111)}$, which is consistent with previous studies in HZO epitaxial films [27].

To confirm the positive correlation between crystallinity and $P_r$, we studied YHO films grown on LSMO(110) / STO(110), which has not been reported in the literature. It turns out, the YHO epitaxial films also grow along the (111) direction, as shown in Fig. 1e. At the optimal growth condition, the YHO(111) / LSMO(110) films also show clear Laue oscillations indicating the sharp surface and interfaces; the spontaneous polarization reaches $\approx 50$ μC/cm$^2$, as displayed in Fig. 1c. The rocking curve of the films displayed in the inset of Fig. 1e is dominated by the sharp peak that represents the large crystal grains, indicating high crystallinity. Combining the data of YHO(111) films grown on LSMO(001) and those on LSMO(110), a clear positive correlation can be observed in Fig. 1f between the crystallinity (large/small grain ratio) and the spontaneous polarization. Overall, the spontaneous polarization increases with the crystallinity and appears to saturate at a value close to 50 μC/cm$^2$.

**Local switching and temperature dependence of polarization**

To cross-check ferroelectricity in the YHO films, samples were measured using piezoresponse force microscopy (PFM) as well as temperature-dependent measurements of the polarization hysteresis loops.

Figures 2a,b show PFM images of the bipolar domain patterns written by an electrically biased tip on the YHO(111) / LSMO(001) film surface. An enhanced PFM amplitude signal was obtained in the electrically poled regions (inside the square marked with blue dashed lines in Fig. 2a). Clear and well-defined domain walls could be seen on the boundary separating the oppositely poled regions. The corresponding PFM phase images in Fig. 2b reveal bright and dark regions with a nearly 180° phase contrast, corresponding to the downward and upward polarization states, respectively. From the uniform contrast in the PFM amplitude and phase images of the unpoled region, it can be deduced that the polarization in the as-grown film was aligned downward. Comparison of the PFM images with the corresponding topography of the scanned region (Fig. 2c) reveal negligible correlation of the domain patterns and topographic features. Clear terraces and steps in the topography image reveal the high quality of the YHO film. Local PFM spectroscopic measurements (Fig. 2d) demonstrate the electrical switchability of the films. Based upon the clear signature of ferroelectricity obtained from the structural and macroscopic electrical measurements, the observed PFM features are most likely related to the intrinsic ferroelectric behavior, which is not obscured by contribution from the extrinsic factors such as charge injection [40].

Figure 3a shows the remanent $P$-$V$ loops for the YHO(111) /LSMO(001) sample grown at the optimal condition, which shows a weak temperature dependence between 20 and 300 K. For the YHO(111) /LSMO(110) sample grown at the optimal condition, the $P_r$ increased from 37 μC/cm$^2$ at 20 K to about 50 μC/cm$^2$ at 300 K as shown in Fig. 3b. Comparison of the $P_r$ as a function of temperature for the two different films is shown in Fig. 3c. The increase of $P_r$ with temperature is opposite to that of the conventional ferroelectric materials since the ferroelectric order is expected to be higher at low temperature. One possibility is the extrinsic contributions to the polarization switching process such as oxygen migration [41,42]. Recently, in-situ STEM was employed to study the ferroelectric switching process in the HZO(111) / LSMO(001) thin film structures, where HZO stands for Hf$_{0.5}$Zr$_{0.5}$O$_2$. It was reported that most of the polarization in the HZO(111) / LSMO(001) structures could be attributed to oxygen vacancy migration,[41] with the intrinsic ferroelectric polarization estimated to be less than 9 µC/cm$^2$. In addition, it was reported that $P_r$ in the 5.6-nm-thick HZO films grown on LSMO/LaNiO$_3$/CeO$_2$/YSZ/Si(100) decreases by a factor of 3 from 300 K to 20 K.[43] On the other hand, for the YHO(111) films studied in this work, $P_r$ exhibits a large value at low temperature and stays nearly constant below 100 K. This suggests that the extrinsic contributions are minimal to the measured $P_r$ at low temperature. The moderate overall change of $P_r$ with temperature also indicates that the contribution from the intrinsic ferroelectricity dominates even at room temperature.

Further evidence for the minimal effect of the oxygen vacancies migration was obtained by comparing the imprint behavior at 300 K (Fig. S6a) and at 20 K (Fig. S6b,c). Recently, it was reported that the imprint in HfO$_2$-based films is strongly dependent on their poling history, i.e., positive (negative) imprint would develop if the last switching pulse would set the capacitor to the upward (downward) polarization state [44]. This so-called fluid imprint could also be observed in our samples at room temperature (Fig. S6a). However, upon cooling to 20 K, the imprint remained 'frozen-in', which can be attributed to the minimal movement of internal charges (such as oxygen vacancies) at low temperatures.

Finally, remanent *P-V* loops were measured at 20 K for the YHO(111) /LSMO(001) samples with different crystallinity, to verify if the trend of increasing $P_r$ with crystallinity was intrinsic in origin. As shown in Fig. 3d, strong correlation between the high $P_r$ and high crystallinity observed at room temperature could be reproduced at 20 K (see also Fig. 1d), suggesting the intrinsic nature of the observed features.

**o-phase with a rhombohedral distortion and the *t→o* structural transition**
The high crystallinity of the YHO films allows for determination of the crystal structure, which is critical for understanding the ferroelectricity. Previously, the ferroelectric HZO(111) / LSMO(001) films have been found to be the *o*-phase [25-27]. Other work identifies a rhombohedral unit cell or possibly a rhombohedral phase in HZO(111) / LSMO(001) [19,24,37]. Here we show that the YHO(111) films grown on both LSMO(110) / STO (110) and LSMO(001) / STO (001) are consistent with the Pca2$_1$ *o*-phase with a rhombohedral distortion.

The Pca2$_1$ *o*-phase structure was confirmed by measuring the lattice constants and lattice distortions.

The lattice constants of the YHO were probed by measuring the spacing of the {200}$_{pc}$ planes. For the YHO(111) / LSMO(001) films, due to the 4-fold rotational symmetry of LSMO (001), the (111) oriented YHO films contain four structural domains rotated by 90° relative to each other along the film normal [37,45], which multiplies the three tilted {200}$_{pc}$ planes (tilt angle χ≈55°) to twelve (see Supplementary Fig. S13a). For the YHO(111) / LSMO(001) films, as shown in Fig. 4a, after averaging the twelve directions, the {200}$_{pc}$ diffraction profiles show two distinct peaks, corresponding to lattice constants of 5.20 ± 0.01 and 5.07 ± 0.01 Å, respectively. Overall, the peak at smaller 2θ has about ½ of the area of the other peak, indicating that one lattice constant (assigned as *a*) is 5.20 Å, while the other two lattice constants (assigned as *b* and *c*) are 5.07 Å (see Table 1), because the structural factors of the three {200}$_{pc}$ planes are similar due to the nearly cubic structure. The substantial difference between $a$ and $\{b, c\}$ but very close value between $b$ and $c$ is consistent with the orthorhombic Pca2$_1$ ferroelectric phases [10,11,46].

For the YHO(111) / LSMO(110) films, the 2-fold rotational symmetry of LSMO (110) surface generates two structural domains rotated by 180° relative to each other along the film normal (Supplementary Fig. S13b). Since the structural domain boundaries are like "built-in" defects of the films which reduces the crystallinity, the less structural domains in the YHO(111) / LSMO(110) films may explain their higher crystallinity compared with that in the YHO(111) / LSMO(001) films. The double-peaks feature has also been observed for the $\{200\}_{pc}$ planes (Fig. 4a), corresponding to lattice constants $a = 5.21 \pm 0.01$ Å and $b \approx c = 5.08 \pm 0.01$ Å.

Besides the lattice constants, what distinguishes the $o$-phase from the $t$-phase and the $m$-phase is the space group symmetry including a two-fold screw axis along the polar ($c$) axis and two glide planes perpendicular to the other two axes ($a$ and $b$). As a result, the $o$-phase has two important distortions (Fig. 4b, c) relative to the $t$-phase in terms of the displacements of the Hf sites, allowing the diffraction of the orthorhombic $\{010\}_o$ and $\{110\}_o$ planes which are absent for the $t$-phase.

As shown in Fig. 4d, unlike that of the $\{200\}_{pc}$ planes, the diffraction of the $\{100\}_{pc}$ planes only show one peak, suggesting that the Hf displacement is only along one of the $\{100\}_{pc}$ directions. Furthermore, the spacing of the observed $\{100\}_{pc}$ planes are $5.07 \pm 0.01$ Å and $5.11 \pm 0.01$ Å for the YHO(111) / LSMO(001) and the YHO(111) / LSMO(110) films, respectively; both are closer to smaller lattice constants $b$ or $c$, which is consistent with the $\{010\}_o$ plane and the Hf displacement in Fig. 4b.

The Hf displacement in Fig. 4c has been verified using the diffraction of the $\{110\}_{pc}$ planes ($\chi \approx 35°$). As shown in Fig. 4g, for both the YHO(111) / LSMO(001) and the YHO(111) / LSMO(110) films, only one diffraction peak appears for the $\{110\}_{pc}$ planes. The measured plane spacings are $3.64 \pm 0.01$ Å and $3.65 \pm 0.01$ Å, for the YHO(111) / LSMO(001) and the YHO(111) / LSMO(110) films respectively, consistent with the $\{110\}_o$ planes and the Hf displacement in Fig. 4c. Notice that the $m$-phase allows diffraction of $\{110\}_m$ and $\{011\}_m$ planes, which is expected to show as double peaks.

Previously, the appearance of the $\{110\}_{pc}$ x-ray diffraction peak was employed to measure the [47,48] $t \to o$ phase transition in YHO films. Here we measured the temperature dependence of the $(1\text{-}10)_{pc}$ diffraction peak using RHEED. As shown in Fig. 4f, both the $(1\text{-}10)_{pc}$ and the $(11\text{-}2)_{pc}$ diffraction intensities appear as weak streaks in the RHEED images at room temperature. As shown in Fig. 4g, at high temperature, the $(1\text{-}10)_{pc}$ diffraction is absent while the $(11\text{-}2)_{pc}$ diffraction peak is present, indicating the $t$ phase. When the film was cooled, the $(1\text{-}10)_{pc}$ peak appears at about 450 °C, while the intensity of the $(11\text{-}2)_{pc}$ diffraction peak also increases, indicating a transition to the $o$ phase, which is consistent with the range of transition temperature 350 to 550 °C found in previous studies on YHO [23,47-49].

The rhombohedral distortion was measured from the difference between the spacing of the (111) plane (plane normal pointing out of plane) and that of the tilted $\{111\}_{pc}$, i.e., $(\text{-}111)_{pc}$, $(1\text{-}11)_{pc}$ and $(11\text{-}1)_{pc}$ planes, following the work on HZO(111) films.[24,37]

Fig. 4h displays the 2θ scans of the tilted $\{111\}_{pc}$ planes ($\chi\approx71°$) of the YHO(111) / LSMO(001) films averaged over the four domains. The diffraction peak positions of all the tilted $\{111\}_{pc}$ planes are approximately the same, and well separated from that of the $(111)_{pc}$ plane parallel to the film surface (see Supplementary Fig. S9, 10 and 11 for more details); this indicates a rhombohedral distortion, i.e., the angles between the base vectors $α = β = γ = 89.59$ +/- $0.04°$, (distortion angle -$0.41$+/-$0.04°$), similar with the result obtained on HZO(111) / LSMO(001) in the literature. [24,37] As shown in Fig. 4i, the YHO(111) / LSMO(110) films also exhibit a rhombohedral distortion with a distortion angle -$0.25$+/-$0.02°$.

**Theoretical modeling**

To explore the effects of the rhombohedral distortion on the structural stability and ferroelectric polarization of the YHO, we performed density-functional theory (DFT) calculations [50]. We started from the orthorhombic Pca2$_1$ unit cell of undoped $HfO_2$ where the interaxial angles were $α = β = γ = 90°$. A small rhombohedral distortion was then introduced by reducing the angles while keeping them equal, i.e., $α = β = γ < 90°$. Throughout the calculations, the experimental values of the lattice parameters were assumed and fixed to be $a = c = 5.07$ Å and $b = 5.20$ Å, and only inner atomic positions were relaxed. Similar calculations were also performed for the 5% Y-doped $HfO_2$, where effects of doping were modelled by the Virtual Crystal Approximation [51] (see Methods for details).

The calculated ferroelectric polarization of the orthorhombic Pca2$_1$ phase of $HfO_2$ is about 50.2 μC/cm$^2$, which is consistent with the previous theoretical studies.[10-14] The polarization is directed along the *c*-axis, as enforced by the symmetry of the crystal, so that the *a*- and *b*- components of polarization are zero. The 5% Y doping slightly reduces the polarization down to about 49.9 μC/cm$^2$, which indicates that Y does not play a decisive intrinsic role in high polarization values observed in our experiments, but rather helps to stabilize the orthorhombic Pca2$_1$ phase of hafnia.

Figure 5a shows the calculated *c*-component of the ferroelectric polarization $P_c$ as a function of angle *α*. With a larger rhombohedral distortion (smaller *α*), the $P_c$ remains large, but slightly reduces. This reduction is just ~ 0.1 μC/cm$^2$ for the degree of distortion relevant to our experiment. At the same time the broken Pca2$_1$ symmetry allows the appearance of non-vanishing *a*- and *b*- components of polarization, $P_a$ and $P_b$. Figure 5b demonstrates that the absolute values of both $P_a$ and $P_b$ increase with decreasing *α*, but the magnitudes are much smaller than $P_c$. Y-doped $HfO_2$ exhibits the same tendency as the pristine $HfO_2$ (compare the red and blue lines in Fig. 5a and b), implying an idle role of the doping in the polarization enhancement.

Overall, our DFT calculations reveal that the rhombohedral distortion observed in our experiments is not intrinsic to the bulk $HfO_2$ and most likely results from the strain imposed by the substrate. In addition, the rhombohedral distortion of the degree observed in our experiments does not much affect the large ferroelectric polarization value of the orthorhombic Pca2$_1$ phase of the pristine and YHO.

Importantly, comparing the results of our DFT calculations and experimental data indicates that we observe the intrinsic ferroelectricity of Y-doped $HfO_2$ at low temperature. Experimentally, the remanent polarization saturates with increasing crystallinity at low temperature values ranging from 32 μC/cm$^2$ to 37 μC/cm$^2$, corresponding to the of YHO(111) grown on LSMO(001) and LSMO(110), respectively (Fig. 1f). These values represent the projection of polarization of YHO pointing along the *c* axis to the out-of-plane (111)$_{pc}$ direction. This implies that the total spontaneous polarization along the *c* axis is in the range from 55 μC/cm$^2$ to 64 μC/cm$^2$, which is in good agreement with the results of our DFT calculations.

**Discussion**

Stabilization of the *o*-phase has been a core issue for the ferroelectricity in the HfO$_2$-based materials. Thermodynamic stabilization requires that the *o*-phase has lower energy than all the other phases, especially the *m*-phase. However, the theoretically predicted critical size is smaller than was observed experimentally [17]. Therefore, kinetic stabilization is necessary. During cooling from the high-temperature *t*-phase, formation of the *o*-phase is favored since its interfacial energy with the *t*-phase is smaller compared with that of the *m*-phase [19,20]. This process is more effective if the initial *t*-phase has high crystallinity because the structural defects lead to ill-defined *o/t* and *m/t* interfaces. Since the high growth temperature of epitaxial thin films enhances the crystallinity of the *t*-phase, it is expected to promote the stabilization of the *o*-phase and allows formation of the *o*-phase with high crystallinity during cooling, which explains the results of this work.

Two factors are critical for enhancing the stability of the *o*-phase YHO studied in this work. The first factor is the large surface area and anisotropic strain of the ultra-thin epitaxial films, as suggested previously by theory [12,15]. Experimentally, no rapid annealing is necessary for growing ferroelectric HfO$_2$-based epitaxial films, indicating more effective kinetic stabilization [19,21-28,47,52]. In this work, a small rhombohedral distortion in YHO suggests a presence of anisotropic strain, which could enhance the stability of the *o*-phase by increasing the energy of the *m*-phase. The second factor is Y doping, which has been shown to reduce the energy of the high-temperature symmetric phase. [48,53,54] More importantly, Y doping appears to have a superior effect on enhancing the *o*-phase stability, which has been highlighted recently by the demonstration of ferroelectricity in YHO of 1-μm-thick films [35,36] and even in bulk [20].

**Summary and Outlook**

We have demonstrated that the spontaneous polarization of the YHO(111) films grown on both LSMO(001) and LSMO(110) increases with improving crystallinity, which is opposite to the common expectation that the stabilization of the ferroelectric Pca2$_1$ structure requires formation of nanograins. The measured spontaneous polarization was found to be higher and having a much weaker temperature dependence than that reported in the previous studies, indicating the dominant contribution from the intrinsic ferroelectricity. The highly crystalline YHO(111) films contained the orthorhombic ferroelectric Pca2$_1$ structure with a small rhombohedral distortion. Overall, our results demonstrate that Y doping and the anisotropic strain in epitaxial films grown at high temperature strongly enhance the stability of the orthorhombic ferroelectric phase and thus offer a viable approach to HfO$_2$-based ferroelectrics with high crystallinity. This marks a milestone in understanding and tuning the intrinsic properties as well as expanding the application potential of the HfO$_2$-based ferroelectric materials.

## Methods

**Sample preparation.** The YHO thin films on La$_{2/3}$Sr$_{1/3}$MnO$_3$ (LSMO) bottom electrodes were grown by pulse laser deposition (PLD) with a wavelength of 248 nm on SrTiO$_3$ (STO) substrates. The base pressure of PLD chamber is around $3\times10^{-7}$ mTorr. Before depositions, the STO substrates were pre-annealed at 650°C for 1 hour in PLD chamber. LSMO layer with thickness of ~25 nm was deposited at 600°C under a 60 mTorr oxygen atmosphere. The ceramic 5% Y-doped HfO$_2$ target was synthesized at 1400°C by solid-state reaction using HfO$_2$ (99.99% purity) and Y$_2$O$_3$ (99.9% purity) powders. The growth temperature from 750°C to 970°C, a repetition rate of 2 Hz and an oxygen pressure of 70 mTorr were employed to grow the YHO films. The typical thickness of YHO films is about 9-11 nm. At the end of deposition, the temperature of the films decreases to room temperature with a cooling rate of 10°C/min under an oxygen pressure of 70 mTorr. The platinum top electrodes with thickness of ~15 nm were deposited ex-situ by PLD using shadow mask in vacuum at room temperature. The diameter of top electrodes is from 75 μm to 400 μm.

**X-ray structural characterization.** The structural characterizations, including XRD $\theta$-$2\theta$ scans, rocking curves and x-ray reflectivity (XRR), were performed by XRD (Rigaku SmartLab Diffractometer) using Cu Kalpha (wavelength ~ 1.54 Å). The IP and OOP d-spacing of the {111} planes and that of the {001} , {002}, and {110} planes in YHO films were measured using an area detector (Bruker-AXS D8 Discover Diffractometer, wavelength ~ 1.54 Å).

**Electrical measurements.** For the measurements of the ferroelectric properties at room temperature, a solid Pt tip (RMN-25PT400B, RockyMountain Nanotechnology) in contact with the platinum top electrode was used to apply the voltage pulses using a Keysight 33621A arbitrary waveform generator while the transient switching currents through the bottom electrode were recorded by a Tektronix TDS 3014B oscilloscope. In all measurements, the bias was applied to the top electrode (diameter from 75 μm to 400 μm) while the LSMO bottom electrode was grounded. The low-temperature measurements with temperature range from 20 K to 300 K are implemented using Cryostat, Sumitomo Cryogenics, and the top electrodes are connected using silver paint and silver wires.

**Scanning probe microscopy.** PFM measurements were carried out using a commercial AFM system (MFP-3D, Asylum Research) using Pt-coated tips (PPP-EFM, Nanosensor) in the resonance tracking mode by applying an ac modulation signal of 0.8 V amplitude and a frequency of ~ 350 kHz. The bias was applied through the conductive tip and the bottom electrode was grounded.

**Electron microscopy.** For the images with the view along YHO [1$\bar{1}$0] (same as LSMO [001]): The TEM foil was prepared by conventional cutting, grinding and polishing followed by a precision ion polishing (Gatan PIPS695 tool). (S)TEM images with EDS mappings were obtained by a high-resolution transmission electron microscope (HRTEM) FEI TALOS-F200X equipped with high-angle annular dark-field (HAADF) detectors and energy dispersive X-ray spectrometer (EDX). For the images with the view along YHO [11$\bar{2}$] (same as LSMO [1$\bar{1}$0]): An electron transparent cross section of sample of HfO$_2$/LSMO thin film on STO substrate was prepared using Helios NanoLab Dual Beam 660 SEM. The cross-section sample was mounted on a copper FIB lift-out grid. The thinning of the cross-section sample was started from the bottom of the sample to avoid damaging the top part of the sample where the thin films were deposited. Thus, sample was tilted 52±7°, 52±5°, 52±3°, 52±1.5° and was thinned from 2 μm to less than 100 nm by ion beam with 15 kV and 0.42 nA, 8 kV and 0.23 nA, 5 kV 80 pA, and 3 kV and 20 pA, respectively. The final polishing was done at 2 kV and 20 pA. The sample was characterized using a FEI Tecnai Osiris S/TEM.

**Density-functional theory (DFT) calculations.** First-principles DFT calculations were performed using the plane-wave pseudopotential method implemented in the Quantum-ESPRESSO package [50]. Generalized gradient approximation (GGA) for the exchange and correlation functional and an energy cutoff of 544 eV were used in the calculations. The atomic relaxations were performed with an 8×8×8 k-point mesh until the Hellmann-Feynman forces on each atom became less than 1.3 meV/Å. A 10×10×10 k-point mesh was used for the subsequent self-consistent calculations. The Berry phase method was applied to calculate the ferroelectric polarization. The effects of Y doping was modelled by Virtual Crystal Approximation (VCA) [51], by simulating each Hf-site with a pseudopotential of fractional valence. To neutralize the charge in the structures, O-sites are also treated by VCA.

## Data availability
The data that support the findings of this study are included in the main text and Supplementary Information.

## Acknowledgements
This work was primarily supported by the National Science Foundation (NSF), Division of Electrical, Communications and Cyber Systems (ECCS) under Grant No. ECCS-1917635. The research was performed in part in the Nebraska Nanoscale Facility: National Nanotechnology Coordinated Infrastructure and the Nebraska Center for Materials and Nanoscience, which are supported by the NSF under Grant No. ECCS- 2025298, and the Nebraska Research Initiative. Y.Z. and H.W. acknowledge the support from NSF (DMR-2016453 and DMR-1565822 for the microscopy effort at Purdue University.

## Author contributions
The thin films were synthesized by Y.Y. with the assistance from X.X. and Hao.W. Structure distortion and symmetry were investigated by Y.Y. and X.X. Time-resolved RHEED was studied by Y.Y. and X.L. The local switching and temperature-dependent polarization were studied and analyzed by P.B. under the supervision of A.G. M.L. carried out the DFT calculations under the supervision of L.T. and E.Y.T. (S)TEM experiments were conducted by Z.A. and Y.Z. under the supervision of J.S. and Hai.W., respectively. The study was conceived by Y.Y., P.B. and X.X. Y.Y., P.B., M.L., E.Y.T., A.G. and X.X. co-wrote the manuscript. All authors discussed results and commented on the manuscript.

## Competing interests
The authors declare no competing interests.

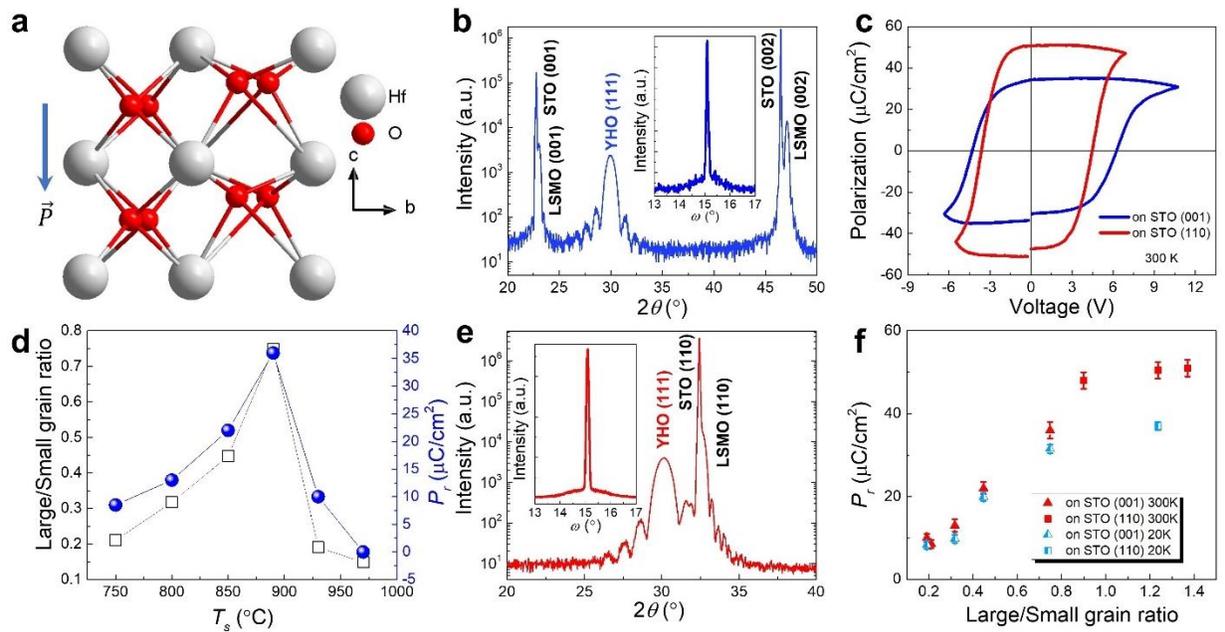

**Fig. 1 | Structure characterization and ferroelectric hysteresis. a,** Crystal structure of $HfO_2$ with $Pca2_1$ space group and a downward polarization. **b,** $\theta$-$2\theta$ XRD scan of YHO(111)/LSMO(001)/STO(001); inset: rocking curve of the YHO(111) peak. **c,** Typical *P-V* loops of YHO(111) / LSMO(001) and YHO(111) / LSMO(110) measured at 300K. **d,** Growth temperature ($T_s$) dependence of the large/small grain ratio (squares) and the remanent polarization ($P_r$) (blue circles) for the YHO/LSMO/STO(001) films. **e,** $\theta$-$2\theta$ XRD scan of YHO(111) / LSMO(110), inset: rocking curve of the YHO(111) peak. **f,** Remanent polarization ($P_r$) as a function of large/small grain ratio at 20K (blue symbols) and 300K (red symbols) on LSMO(111) (triangles) and on LSMO (squares).

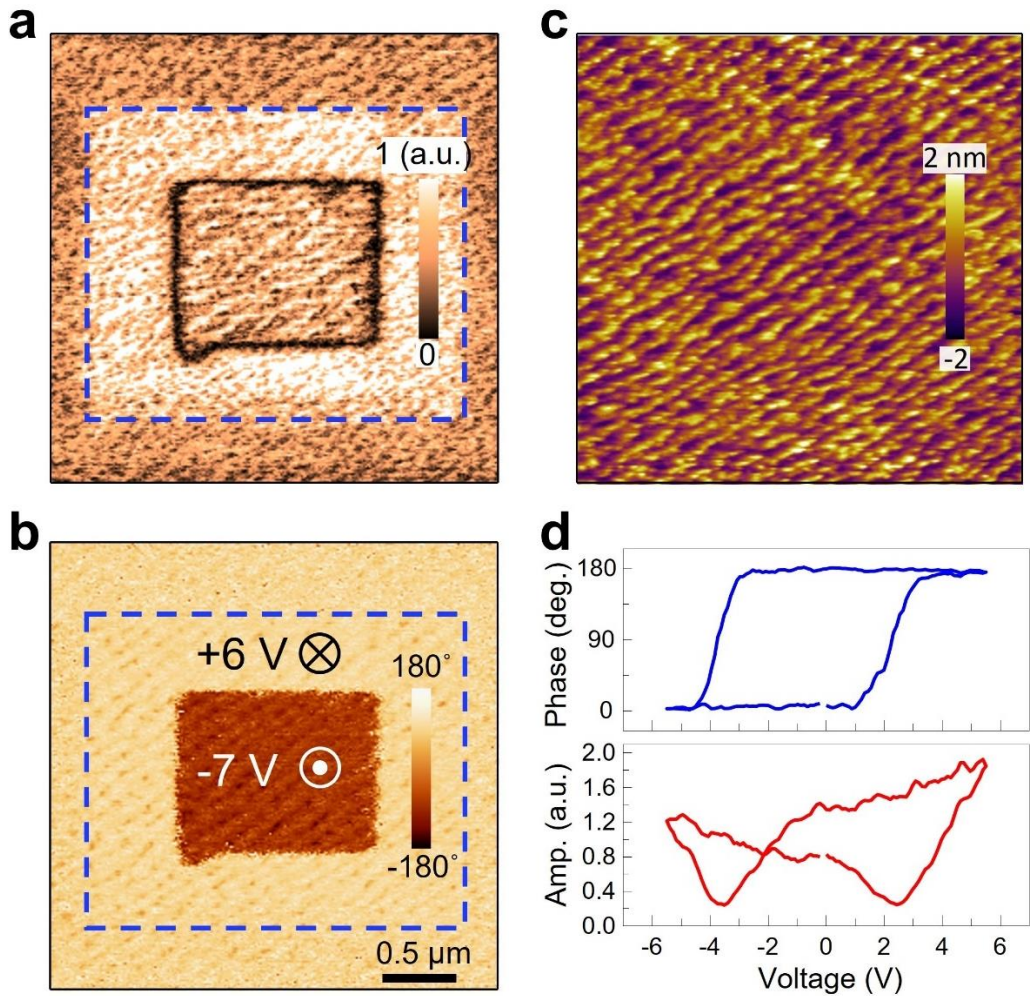

**Fig. 2 | Local ferroelectric switching by PFM. a,** Magnitude and **b,** phase of the PFM image after poling with +6V and -7V, demonstrating stable, bipolar, remanent polarization states. **c,** AFM image of the surface of the YHO(111) / LSMO(001) film, displaying the atomic step-and-terrace morphology. **d,** Phase and amplitude switching spectroscopy loops, demonstrating ferroelectric-like hysteresis on bare surface.

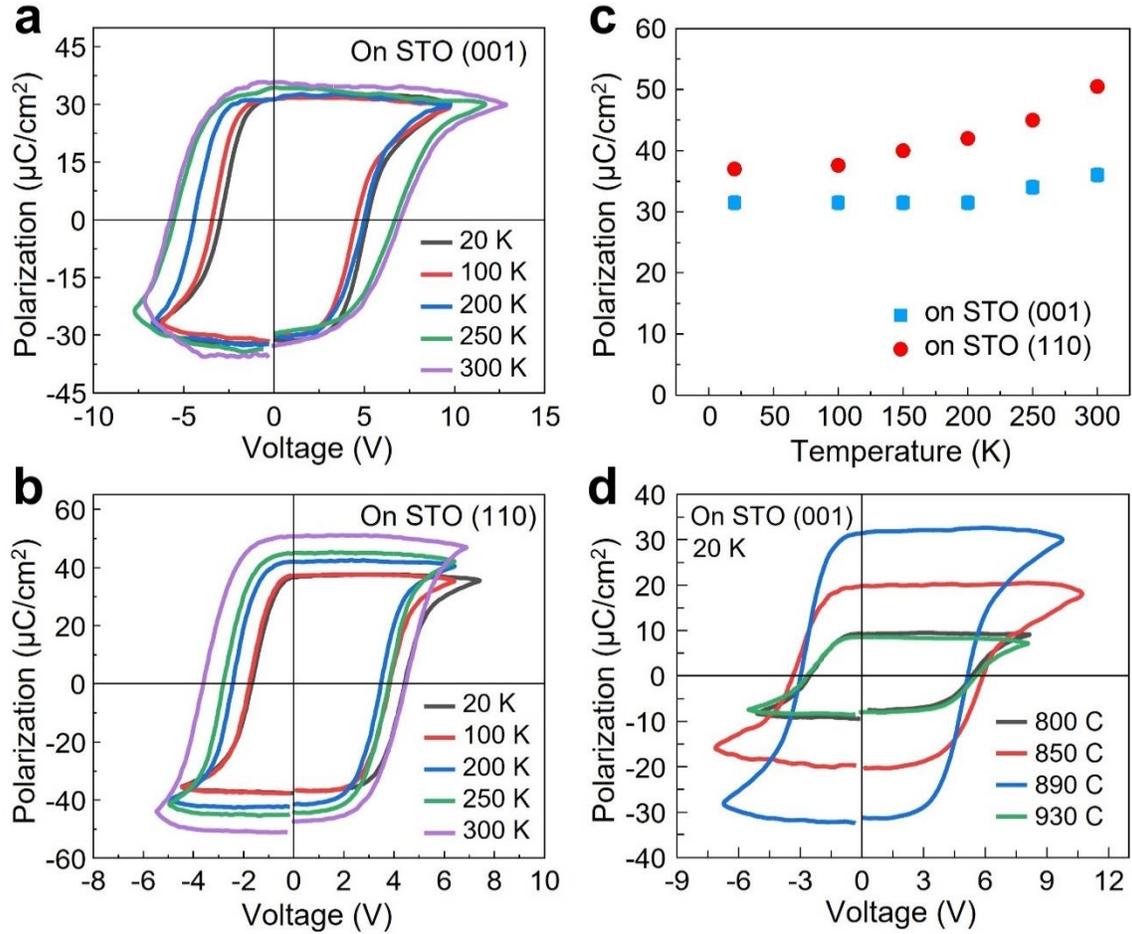

**Fig. 3 | Temperature dependence of the ferroelectric hysteresis.** Ferroelectric *P-V* loops measured by the PUND methods at temperature range from 20K to 300K for **a,** YHO(111) / LSMO(001) and **b,** YHO(111) / LSMO(110). **c,** Remanent polarization as a function of temperature for YHO(111) / LSMO(001) and YHO(111) / LSMO(110). **d,** *P-V* loops for the samples of YHO(111) / LSMO(100) with various growth temperature ($T_s$) measured at 20K.

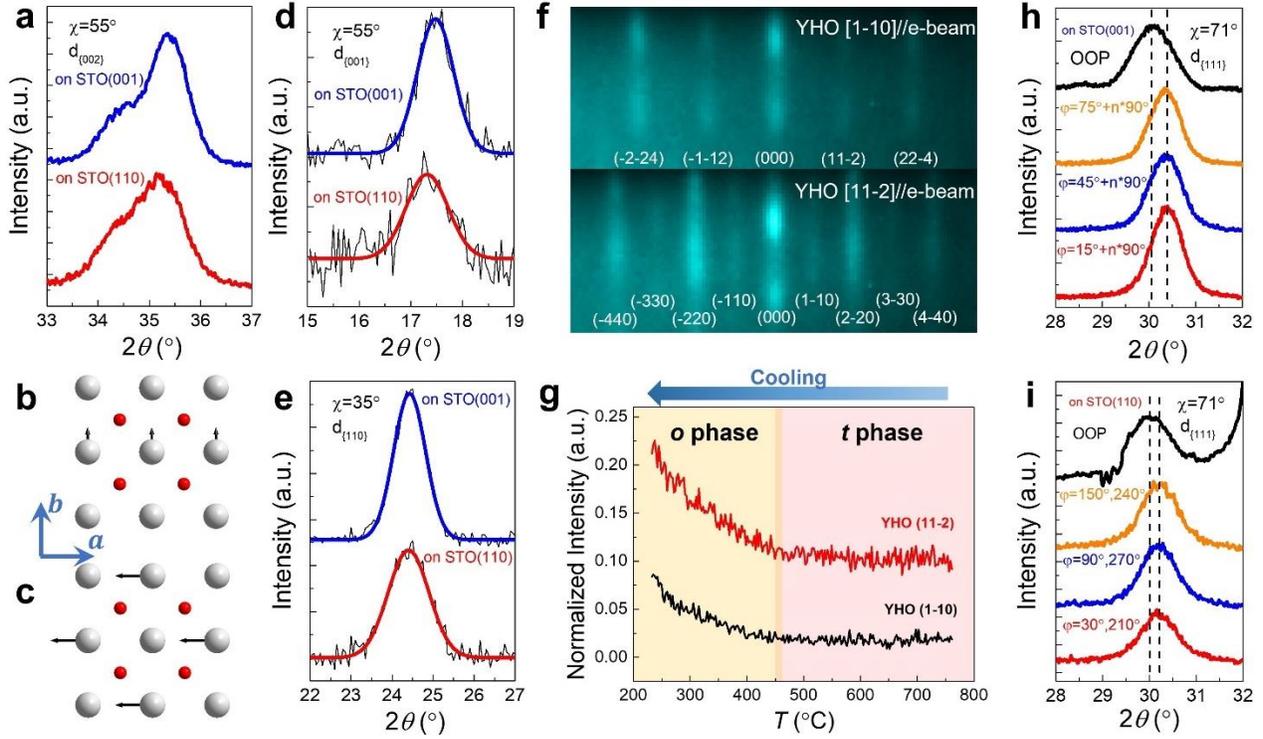

**Fig. 4 | Structural characterization of the YHO films. a,** The diffraction of the {002}$_{pc}$, **d,** {001}$_{pc}$ and **e,** {110}$_{pc}$ plane with χ=55°, χ=55° and χ=35°, respectively. **b,** and **c,** are the displacement patterns of Hf atoms of the Pca2$_1$ *o*-phase HfO$_2$. **f,** RHEED pattern of YHO(111) grown on LSMO(110) along the [1-10]$_{pc}$ and [11-2]$_{pc}$ directions at room temperature. **g,** Temperature dependence of the RHEED intensity of the (1-10)$_{pc}$ and the (11-2)$_{pc}$ streaks, demonstrating a *t→o* phase transition at about 450°C. In-plane and out-of-plane {111}$_{pc}$ diffractions for **a,** YHO(111) / LSMO(001) and **b,** YHO(111) / LSMO(110).

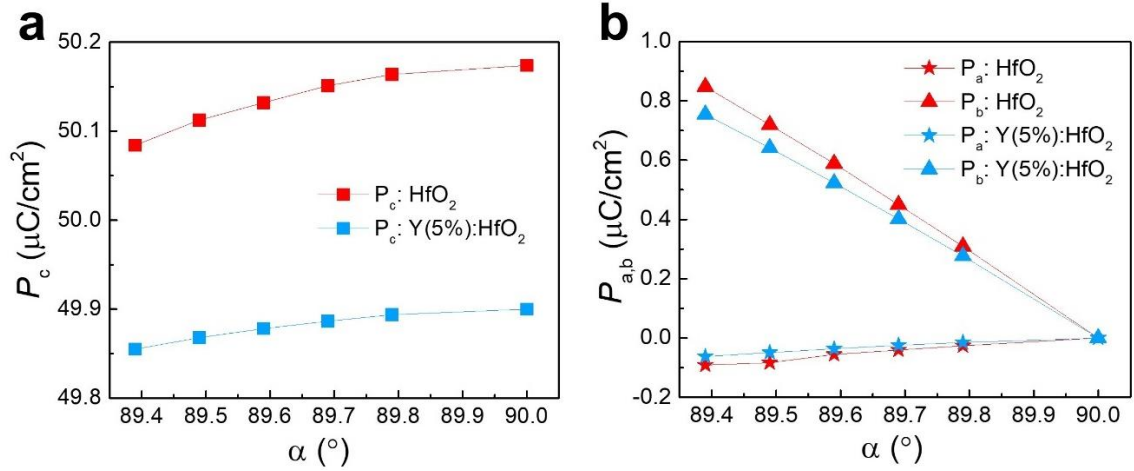

**Fig. 5 | Results of DFT calculations for bulk undoped and 5% Y-doped HfO₂.** Calculated components of the ferroelectric polarization **a,** along the *c*-axis (squares) and **b,** along the *a*- (stars) and *b*- (triangles) axes as a function of angle *α* for bulk undoped HfO₂ (red lines and symbols) and 5% Y-doped HfO₂ (blue lines and symbols).

**Table 1** Structure parameter of YHO films grown on LSMO/STO(001) and LSMO/STO(110) measured by XRD at room temperature.

| Substrate | $a$ (Å) | $b, c$ (Å) | Rhombohedral distortion (°) | Epitaxial relation |
|---|---|---|---|---|
| LSMO (001) | 5.20 ± 0.01 | 5.07 ± 0.01 | -0.41 ± 0.04 | YHO(111) ∥ LSMO(001)<br>YHO[1-10] ∥ LSMO[1-10] |
| LSMO (110) | 5.21 ± 0.01 | 5.08 ± 0.01 | -0.25 ± 0.02 | YHO(111) ∥ LSMO(110)<br>YHO[1-10] ∥ LSMO[001] |